\documentclass[12pt]{iopart}

\usepackage{graphicx}
\usepackage{iopams}
\usepackage{setstack}
\usepackage[justification=centering]{caption}
\begin{document}

\title{About electrons and position in Triplet Production: some remarks}

\author{ G. O. Depaola$^1$, M. L. Iparraguirre$^1$, D. Palacios$^2$}

\address{$^1$Facultad de Matem\'{a}tica, Astronom\'{\i}a y
F\'{\i}sica, Universidad Nacional de C\'{o}rdoba. Dr. Medina Allende
s/n. Ciudad Universitaria. C\'{o}rdoba 5008. Argentina.}
\address{$^2$Instituto Universitario Aeron\'{a}utico. Av. Fuera Aerea.C\'{o}rdoba 5000. Argentina. }

\ead{depaola@famaf.unc.edu.ar}
\vspace{10pt}

\begin{abstract}
Taking into account the increasing interest in measuring high energy gamma ray polarization, Boldishev et. at. \cite{[Boldy]} published an extensive and very comprehensive work on the possibility of using the recoil electrons in the production of pairs on electrons. However, this work is based on using only 2 Feynmann diagrams of the 8 that the process has. This eliminates the difficulty of distinguishing, in the theory, which is the recoil electron and which is the created

In this work we have analyzed the eight Feynman diagrams and we have shown that for energies lower 
to $\sim 1000mc^2$, the assumption just described is not a good approximation, so we propose a 
different way to work \cite{Marcos}: we classify the electrons into the less energetic and the most energetic ones 
without taking into account their origin.
Under these conditions (lower or higher energy value), we have calculated the contribution of the 
different diagrams to the distribution(we compare the sum of them with that obtained by Haug \cite{Haug_e+}\cite{Haug_e-}, and how these distributions are modified by introducing a 
threshold for the momentum detection for electrons.

For the study of polarization we presented on the angular distribution of particles for 
high-energy gamma rays 
(where only Borsellino diagrams predominate). 
Our results on the azimuthal distribution show that it is highly influenced by the orientation (in 
the plane perpendicular to the direction of the photon), prior to the interaction, that the 
polarization vector has with respect to the position of the electron in whose field the pair will 
be generated. 
\end{abstract}

%
%
%
%
%

\section{Introduction}

\begin{figure}
  \centering
  \includegraphics[scale=0.5]{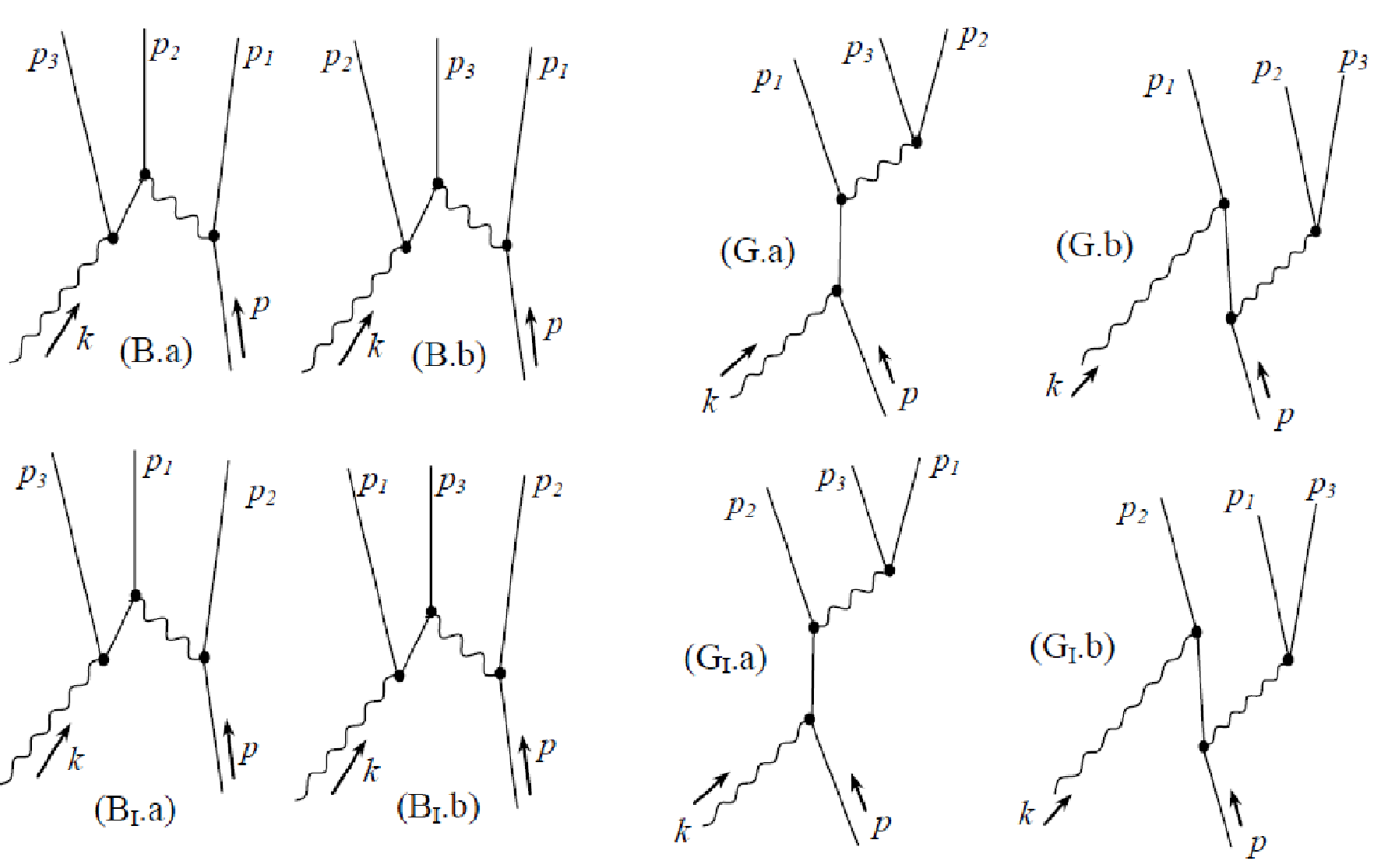}\\
  \caption{Feynman diagrams describing photoproduction of $e^{+}$ $e^{-}$
  pairs on free electrons.}\label{Figure 1}
\end{figure}

In the present days, there are several proposals for the development of instruments to measure 
polarization of high energy gamma rays of extraterrestrial origin. The changes measured 
by INTEGRAL in the polarization direction of the CRAB nebula (\cite{Moran}) in the optical range 
and low energy gamma rays demonstrate that the study of this characteristic can provide much 
relevant information. 

When a high-energy gamma ray interacts with matter, it is most likely to interact with the electric field of either the atomic nucleus or an electron of the atom and become an electron-positron pair. When it does so with the atomic nucleus, given the high mass of the nucleus (with respect to the $e^{-} - e^{+}$ mass) the energy absorbed by the nucleus is considered negligible (but not the momentum) and it is assumed that all the energy of the gamma ray is transmitted to the created pair (Bethe-Heitler representation). If the interaction takes place with the electric field of an electron, the energy absorbed by it can be sufficiently high not to be neglected and therefore, as a final result of the process, in addition to the created pair, there is an extra electron (called the recoil electron) constituting what is called a triplet. The probability ratio between the two cases is $1/Z$, therefore one is more likely to observe triples in low $Z$ atoms.
Given the high energy of the gamma rays that are usually converted into pairs, the pair created have a very high probability of having very small opening angles (relative to the direction of the gamma ray, polar angle) and this is the same for either of the two processes already mentioned. If the beam is linearly polarized, the direction of the polarization vector is reflected in an asymmetry of the probability distribution of the pair around the axis defined by the direction of the gamma ray (azimuthal angle). This asymmetry is generally small and depends on how far the plane defined by the momentum of the pair moves away from the direction of the gamma ray (coplanarity). This makes the measurement of the asymmetry very dependent on the ability to measure the angles involved (\cite{depa_astro}, \cite{depa_rpc}, \cite{Maximon}) and it becomes much more difficult for high-energy gamma rays because the polar angles are very small and multiple scattering tends to destroy this information.

In the case of the triplet, it is most probable that one of the three particles has a large polar angle and, with a little less probability, this particle will possess sufficient energy to be detected using gas detectors ( \cite{harpo}, \cite{hunter}) or Si detectors (\cite{wojt} \cite{dugger}, 
\cite{astrogaam}) or emulsion technique, \cite{emulsion} making the measurement of polar and azimuthal angles more easier.

The study and analysis of the angular distribution of the electron recoil promises to be an effective method to help to determine the polarization of the gamma ray along with the study of the pair but the indistinguishability of the electrons makes it difficult to identify which is the created electron and which is the electron recoil. 

In order to understand and extract the desired parameters, it is necessary to study how this 
indistinguishability of electrons affects the experimental data.

The momentum (or energy) distributions for the three particles can be calculated by using the 
standard technique of the quantum electrodynamics in the lower approximation order. For this 
process:

\begin{equation*}
    \gamma + e^{-} = e^{+} + e^{-} + e^{-}.
\end{equation*}

\noindent 
one has 8 Feynman diagrams, two of them, called the Borsellino diagrams \cite{[Borse]} (Ba and Bb 
in fig. \ref{Figure 1}), make the main contribution to the total cross section of the process 
(\cite{[JosephRohrlich]}, \cite{[Mork]} \cite{[Boldy]}). Joseph and Rohrlich \cite{[JosephRohrlich]} also show that the process in witch the final state configuration consist with the positron and one electron emerging with very high energy is by far the dominant contribution to the cross section.

Many authors put focus only in these 
diagrams to obtain the momentum or angular distributions of the particle ( Boldyshev et. al. 
\cite{[Boldy]} made an extensive calculation, for example).

The use of only these two diagrams simplifies the interpretation of the role of the electrons 
since, according to figure \ref{Figure 1}, it is clear that $p_{1}$ distribution corresponds to the 
recoil electron and $p_{2}$ to the created electron. This is a good approximation for high energy 
limits that we have established in this work as $\omega \geq 1000m$. This facility is lost when 
we include the exchange diagrams.

In this work we have shown that for medium and lower energies, the contribution of the other 
diagrams, in particular the inclusion of the exchange of the Borsellino diagrams overestimate the 
total cross section making it necessary to take into account all the diagrams (this was also quote by Haug \cite{Haug_e+}). Also, the probability 
that the created electron could acquire energy lower than the recoil electron can be significant in 
the medium and lower regime, and the inclusion of the exchange diagrams change the role of these 
particles. This is important since in an experiment it is not possible to distinguish between the 
pair electron and the recoil electron. Our proposal is to calculate the energy distribution in a 
situation more similar to the experimental setup, that is, to calculate the energy distribution of 
the lower energetic electron and the energy distribution of the higher energetic electron without 
taking into account the origin of the electrons.

In section 2, we have given a brief discussion about how we have calculated the cross section from 
the Feynman diagrams, which variables we have chosen to integrate analytically and which ones, in 
numerical form, and how to find the limits that the kinematics has imposed on the variables of 
interest. The most important diagrams are the Borsellino one and the respective exchange where the 
labels of the electrons are inverted. We concentrate on the calculations based on these diagrams for explanation.

The section 3 shows the energy distribution given by the different terms of the scattering matrix. We have also tried to show where only the use of the 
Borsellino diagrams is a good approximation analyzing this overlapping of the distribution that 
quantifies, in some sense, the probability the role is exchanged between the electrons. In section 
3a, we have calculated the energy distribution obtained for the lower energetic electron and for the 
higher energetic electron in 3b. We have also shown the contribution for many terms of the 
scattering matrix and compared the sum with the closed formula obtained by Haug \cite{Haug_e-}.

Angular distribution is very important for determining polarization. This is discussed in section 4 
for high-energy gamma rays (only Borsellino diagrams were used). It starts from the triple 
differential cross section and analyzes the distribution of the polar (section 4) and azimuthal 
angle (section 5) for 
the different particles and in different regimes of their momenta.

\section{Feynman diagrams and some kinematics considerations}

The 4-momentum conservation for this collision is written as:
\begin{equation*}
    k + p = p^{-}_{r} + p^{-} + p^{+} = p_{1} + p_{2} + p_{3}.
\end{equation*}

As usual, in this work we take $\hbar $ and \textit{c} equal to 1, and then, in the laboratory 
system, \textit{k} = ($\omega$, $\omega \hat{k}$), and p = (m,$\vec{0}$), where 
$\hat{k}$ is the unit vector of the incidence $\gamma$ - ray direction . We call $p_{1}$ and $p_{2}$ 
to the 4-momentum of the electrons that, depending on the diagrams, (figure \ref{Figure 1}) can be 
the recoil electron or the momentum of the created electron or conversely. $p_{3}$ is the 4- 
momentum of the positron (we use $\rm p_i$ = $|\vec{\rm p_i}|$ for the 3-momentum module).

The cross section of this process can be calculated by using the standard technique. The 8 Feynman 
diagrams are shown in figure \ref{Figure 1}, 2 of them, the $B.a$, $B.b$ are the Borsellino 
diagrams, the other 2, $G.a$ and $G.b$, are called $\gamma - e^{-}$ and the other 4 ($B_{I}.a$, 
$B_{I}.b$, $G_{I}.a$, $G_{I}.b$) are the exchange diagrams due to the indistinguishability between 
the recoil electron and the electron created (in this last 4th case, the role of $p_{1}$ and $p_{2}$ 
interchanges). 

Looking at the $G$ diagrams, one can see that these are basically Compton diagrams, with a created 
pair at the end of the virtual photon, so it is expected that the contributions to the cross section 
of these diagrams will decrease with the gamma-ray energy.

To illustrate the calculation, we have neglected the $G$ diagrams (however, in our final energy distributuion, we 
have included all combinations making a total of 10 matrix elements to add ). Under these conditions, the matrix elements to be calculated are: 

\begin{equation}
 |M|^2  =  |M_{B} - M_{BI}|^2 
 =  |M_{B}|^2 +|M_{BI}|^2 - (M_{B} M^*_{BI}- M^*_{B} M_{BI}) 
\end{equation}

The first effect to notice with respect to the usual calculations using only the Borsellino 
diagrams, is the introduction of the term $|M_{BI}|^2$ which consists  in the exchange of the label 
between the two electrons, that is $|M_{B}(1,2)|^2 +|M_{BI}(2,1)|^2$.

The second and the most important contribution is the interference term which give how much 
overlapping is, in the phase space, between the variables; for example, once one integrate over all 
the variables except for $p_{1}$,

\begin{equation}
 \frac{d\sigma}{d^3 \rm p_{1}} = \int\int\int |M|^2 \cdots d^3 \rm p_{2} d^3 \rm p_{3}
\end{equation} 

\noindent 
one will have the probability density that the recoil electron and the created electron can 
take the value $\rm p_{1}$. 

The inclusion of the exchange term has made it necessary to multiply the cross section by the $1/2$ 
factor to avoid double-counting due to the indistinguishability of the electrons.

The triple differential cross section for the process is:

\begin{equation}
d\sigma = \frac{\alpha r^2_0}{4 \pi^2}\frac{1}{\omega m}\frac{d^3 \overrightarrow{\rm p_1}}{E_1}\frac{d^3 \overrightarrow{\rm p_2}}{E_2}\frac{d^3 \overrightarrow{\rm p_3}}{E_3}\delta^4\left( \sum p_f - \sum p_i\right) \left( \frac{4 m^6}{e^6} |M|^2\right) 
\end{equation}

Since in this work we are interested to describe the electrons distributions, we have used the 
delta function to integrate over the positron variables ($\overrightarrow{\rm p_3}$),

\begin{equation}
d\sigma =  \frac{\alpha r^2_0}{2 \pi}\frac{1}{\omega m}\frac{d^3 \overrightarrow{\rm p_1}}{E_1}\frac{d^3 \overrightarrow{\rm p_2}}{E_2}\frac{1}{E_3(\overrightarrow{\rm p_1}, \overrightarrow{\rm p_2})} \\
 \delta\left( \sum E_f - \sum E_i\right) \left( \frac{4 m^6}{e^6} |M|^2\right)
\end{equation}

\noindent 
where in the $M$ we  also must substitute $E_3$ and $\overrightarrow{\rm p_3}$ by the expressions 
as a function of $\overrightarrow{\rm p_1}$ and $\overrightarrow{\rm p_2}$ using the conservation law.

\begin{figure}
  \centering
  \includegraphics[scale=0.6]{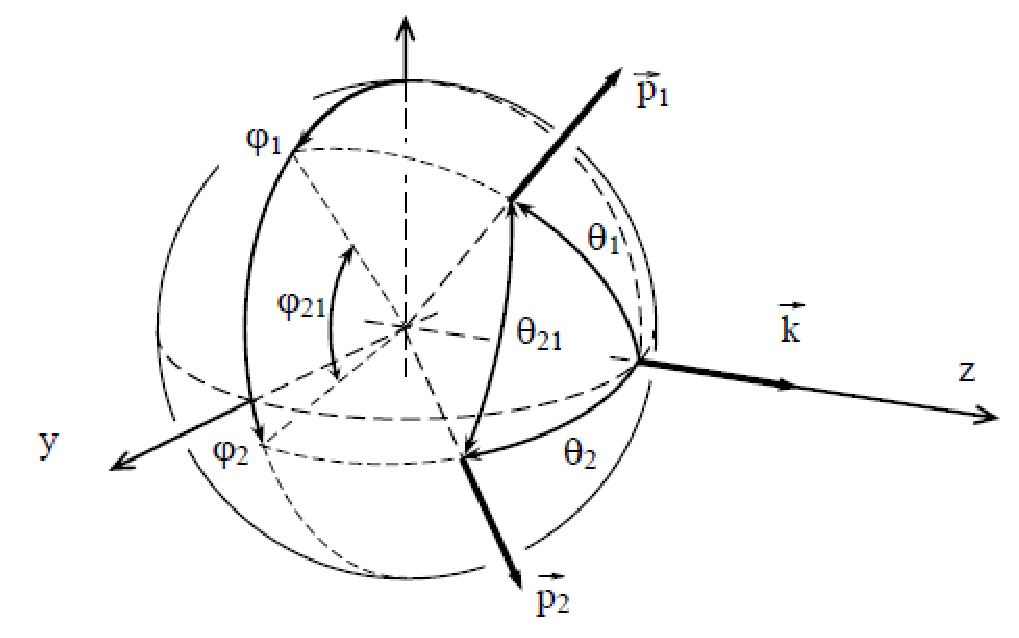}\\
  \caption{Angles after integrate over positron variables}\label{Figure 2}
\end{figure}

The last delta function has been chosen to integrate over the azimuthal angle of electron labeled 2, 
$\varphi_2$:

\begin{equation}
d\sigma =  \frac{\alpha r^2_0}{2 \pi}\frac{1}{\omega m} \frac{d^3 \overrightarrow{\rm p_1}}{E_1}\frac{\rm p^2_2 \sin \theta_2 d \rm p_2 d \theta_2}{E_2 (2 \pi \rm p_1\rm p_2 \sin \theta_1 \sin \theta_2 |\sin(\varphi_2 - \varphi1))} \\
 \left( \frac{4 m^6}{e^6} |M|^2\right) 
\end{equation}

The absolute value of the expression corresponds to, once $\varphi_1$ is fixed, two possible values 
to $\varphi_2$ that are given by the same $\theta_{21}$ (see figure \ref{Figure 2}). The $\varphi_2$ 
value must void the following expression (where $x$ represents the other variables):

\begin{equation}
 g(x,\varphi_2) = \omega + m -E_1 - E_2 - E_3(\overrightarrow{\rm p_1},\overrightarrow{\rm p_2})
\end{equation}

The condition $g(x, \varphi_2)=0$ gives the following expression for the $\cos \varphi_{21}$ where 
$\varphi_{21} = |\varphi_{2} -\varphi_{1}|$:

\begin{equation*}
 \cos \varphi_{21}  =  \frac{1}{\rm p_1 \rm p_2 \sin \theta_1 \sin \theta_2}\left[ \left( \omega + m\right) \left( m - E_1 - E_2\right) + \right.
\end{equation*}
 
\begin{equation}
\\ \left. E_1 E_2 + \omega ( \rm p_1 \cos \theta_1 + \rm p_2 \cos \theta_2)- \rm p_1 \rm p_2 \cos \theta_1 \cos \theta_2 \right] 
\end{equation}

This expression is not necessarily less than $1$ when $\overrightarrow{\rm p_1}$ and $\overrightarrow{\rm 
p_2}$ are into their validity range given by the kinematics (see figure 3  in ref \cite{[Ger]}). 
Once fixed $\overrightarrow{\rm p_1}$, this expression is very restrictive to $\overrightarrow{\rm 
p_2}$ \cite{Marcos}. Defining $\rm F$ as:

\begin{equation}
 \rm F =  - \rm p^2_2 \cos^2 \theta_2 G + 2 \rm p_1 \cos \theta_2 (\omega - \rm p_1 \cos \theta_1) C + 
 \rm p^2_1 \rm p^2_2 \sin^2 \theta_1 - C^2
\end{equation}

\noindent 
where:

\begin{equation*}
 \rm C = (\omega + m) ( E_1 + E_2 - m)  E_1 E_2 - \omega \rm p_1 \cos \theta_1
\end{equation*}

\begin{equation*}
 \rm G = \omega^2 + \rm p^2_2 - 2  \omega \rm p_1 \cos \theta_1)
\end{equation*}

\noindent 
the differential cross section can be written as: 

\begin{equation}\label{dsima}
d\sigma =  \frac{\alpha r^2_0}{2 \pi}\frac{1}{\omega m} \frac{d^3 \overrightarrow{\rm p_1}}{E_1}\frac{\rm p^2_2 \sin \theta_2 d \rm p_2 d \theta_2}{E_2} 
 \frac{1}{\pi \sqrt{\rm F} } \left( \frac{4 m^6}{e^6} |M|^2\right) 
\end{equation}

\noindent  
where the extreme values of $\theta_2$ (once $\overrightarrow{\rm p_1}$ and $\rm p_2$ are into 
their validity range) must satisfy the condition $\rm F\geqslant 0$

\begin{equation}
\cos \theta^{\pm}_2 = \frac{(\omega - \rm p_1\cos \theta_1) C \pm \rm p_1 \sin \theta_1 \sqrt{\rm p^2_2 G - C^2}}{p_2 \rm G}
\end{equation}

\noindent 
(it has been found that the extreme values for $\theta_2$ are very close each other).

The differential cross section (eq.\ref{dsima}) admits two analytical integration, one trivial over 
$\varphi_1$ since in any stage of the calculation one can write the cross section as $d \sigma = d 
\sigma^t - d \sigma^l \cos 2 \varphi_1$, the other chosen variable was $\rm \theta_2$ (we have omitted 
the final expression because it is too long), so finally we have an exact expression for the 
differential cross section according to $d \sigma = d \sigma(\rm p_1, \rm p_2, \theta_1)$. To obtain 
the momentum distribution we must continue with numerical integration.

Haug, using another integrating technique, obtained $d \sigma = d \sigma(\rm p_3,  \theta_3)$ \cite{Haug_e+} and $d \sigma = d \sigma(\rm p_2, \theta_2)$ \cite{Haug_e-} where $\rm p_2 $ and $\rm \theta_2$ are the momentum and polar angle of the electrons (created and recoil), that is, the sum of the each contribution.

\begin{figure}
  \centering
  \includegraphics[scale=0.8]{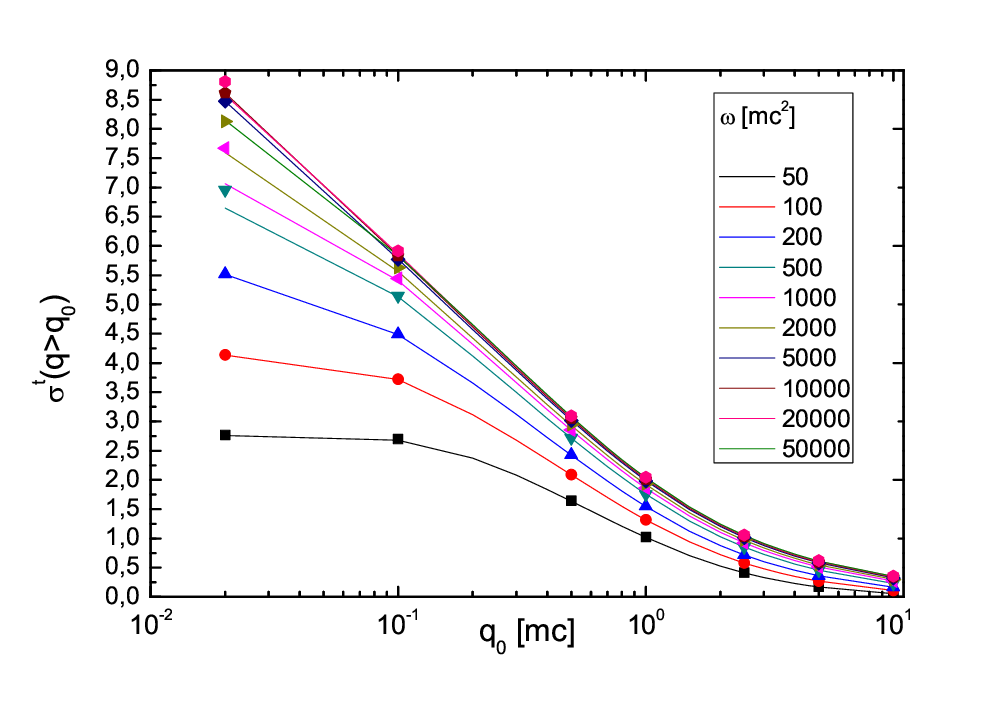}\\
  \caption{Comparison of total cross section as a function of threshold momentum for many gamma rays 
energies (lower energy at bottom) between our calculation (line) and table V in \cite{[Boldy]} 
(symbols)}\label{Figure ber}
\end{figure}

In order to test our calculation, in figure \ref{Figure ber} we compare the total cross section in 
function of the threshold momentum (only for the recoil electron) obtained in this work (lines) with 
the table V in \cite{[Boldy]} (symbols) obtained using only the Borsellino diagram in both cases. 
One can see that for $\omega \geq 5000 m$ the curves can not distinguish the asymptotic 
expression \cite{[Ger]} when $q_0 \geq 0.1 m$. Bernard \cite{[Bern]} also produces these curves 
including all the terms (figure 6 in \cite{[Bern]}) but it is noticed that the values obtained are 
lower respect to Boldyshev and this work, probably because all the terms have been used (and in this 
case it is not clear how the threshold is applied).

\section{Energy distribution for the electrons}

To obtain the energy distribution of $\rm p_1$ that, as one can see from the Feynman diagrams in 
fig.1, which represents the recoil electron in $|M_B|^2$ and the created electron in $|M_{BI}|^2$, 
we have integrate (in numerical form) over $0\leq \theta_1 \leq \theta_{1 max} = \arccos\left( 
\frac{\omega (E_1 - m) + m (E_1+m)}{\omega \rm p_1}\right)$ (the maximum $\theta$ allowed by the 
kinematics, eq. 28 in \cite{[Boldy]}) and over $\rm p_{2min}(\theta_1, \rm E_1) \leq \rm p_2 \leq \rm 
p_{2max}(\theta_1, \rm E_1)$ (eq.7.3 in \cite{Haug_e+}), roots of $\rm p_2^2G=C^2$ (see figure 3 in 
\cite{[Ger]}).
Another fact to take into account is that in an experiment there is a threshold for the momentum 
detection below where the particle is not detected. In our calculations, we have designated to this 
limit as $q_0$ and if $\rm p_{min} \leq  q_0$ the integration is done between $\rm q_0$ and $\rm 
p_{max}$.

\begin{figure}
  \centering
  \includegraphics[scale=0.6]{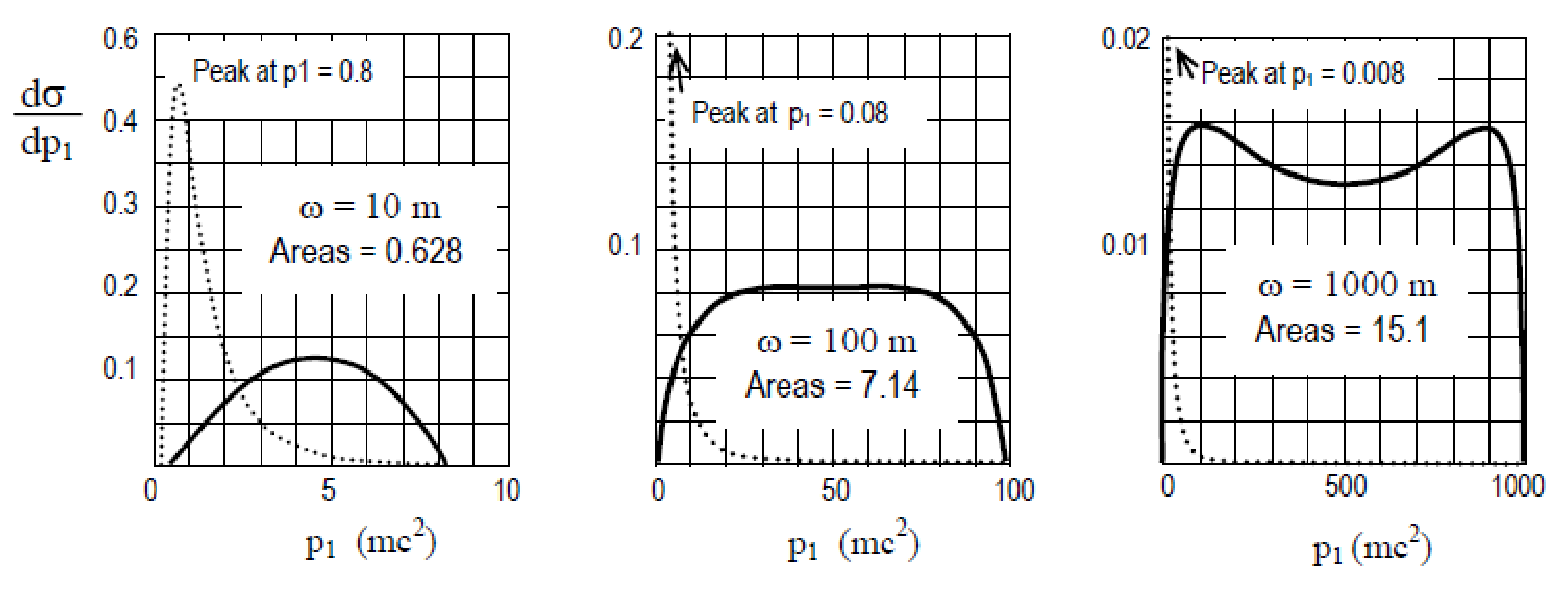}\\
  \caption{Energy distribution for $\rm p_1$ obtained using only $|M_B|^2$ where represent the 
recoil electron and the energy distribution obtained using $|M_{BI}|^2$ only, where $\rm p_1$ is the 
electron created in this diagrams.}\label{Figure 3}
\end{figure}

In figure \ref{Figure 3} we show the energy distribution for $\rm p_1$ obtained using only 
$|M_B|^2$ where it represents the recoil electron and the energy distribution obtained using 
$|M_{BI}|^2$, where $\rm p_1$ is the distribution for the electron created in this case. We show the 
results for 3 different $\omega$, $10 m$, $100 m$ and $1000 m$. As it was mentioned before, the two 
curves have the same area (shown in the figure and in agreement with that one obtained by Joseph and 
Rohrlich \cite{[JosephRohrlich]}), the $|M_B|^2$ was truncated to see $|M_{BI}|^2$. The peak of the 
$|M_B|^2$ is found when $\rm p_1 \backsimeq 8/\omega$. If the value for the threshold detection for 
$\rm p_1$ is around $ 1 m$, in general, only one sees a decreasing distribution for the electron 
recoil.

From this figure, one can deduce that for $\omega \lesssim 1000 m$  the probability that the recoil 
electron has acquired energy greater than the created electron, is not so small, so to compare the 
theoretical distribution with the experimental data we propose to calculate $d \sigma_T (\rm p_1 < 
\rm p_2)/d \rm p_1$  and $d \sigma_{T} (\rm p_1 > \rm p_2)/d \rm p_1$ ($d \sigma_{T}$ is the sum of 
all 10 terms) obtaining in this way the energy distribution for the lower and higher energetic electron 
whatever their origin is. 

\subsection{Energy distribution for the lower energetic electron}

\begin{figure}
  \centering
   \includegraphics[scale=0.8]{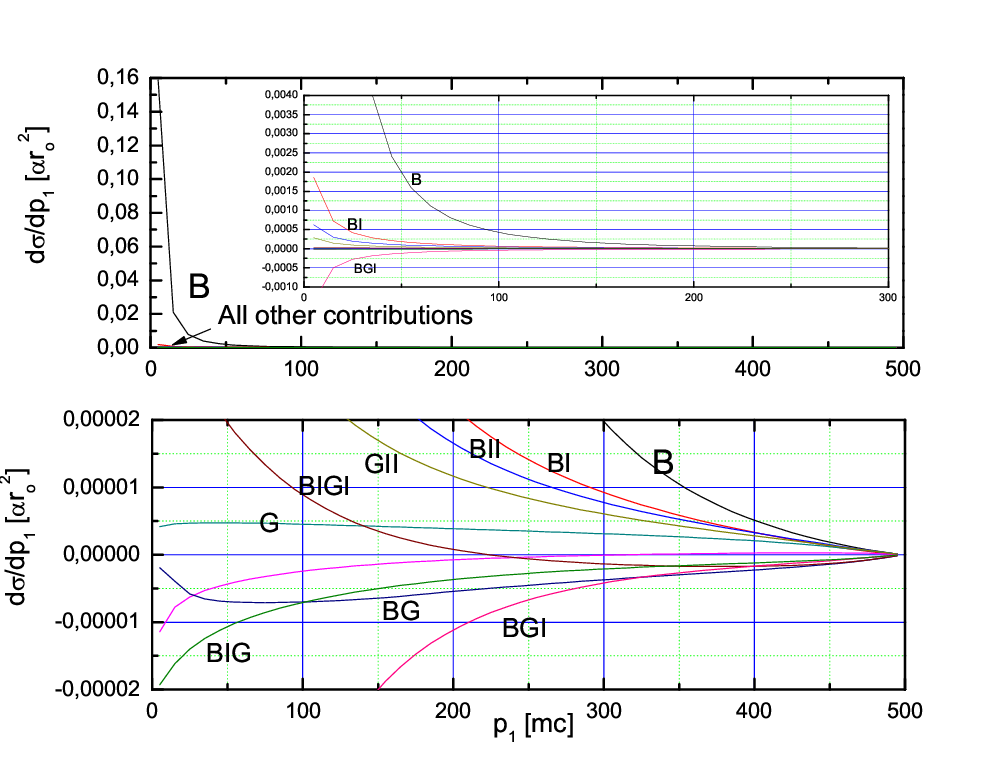}\\
  \caption{Momentum distribution for the lower energetic electron using different contribution of the 
Feynman diagram for $\omega=1000m$.}\label{Figure 4}
\end{figure}

In the figure \ref{Figure 4} we show, for $\omega=1000m$, the momentum distribution for the lower 
energetic electron using different contribution of the Feynman diagrams. The symbology used is the 
following one: $B$ represents the contribution of $|M_B|^2$, $BII$ for $|M_{BI}|^2$, $BI$ $\equiv$ 
$|M_{BI} M^{*}_{BI}-M^{*}_{BI} M_{BI}|^2$, $G$ $\equiv$ $|M_{\gamma e}|^2$ and so on ($GI, GII, BIG, BGI, BG, BIGI$). Figure \ref{Figure 4}a 
shows the full scale and one can see how dominant is the $|M_B|^2$ contribution. Figure \ref{Figure 4}b is 
the same as figure 4a but with a zoom to see the other contributions. In table 1 we compare the 
contribution of each term in relation to the areas of each curve respect to the area of the 
Borsellino term (in percentage) for three gamma-ray energies. It is noteworthy that the only 
inclusion of the Borsellino exchange diagram imply a correction of 8.15 \% for $100 m$ and this 
overestimates the correction that includes all terms, which is 5.88 \%. 

\begin{table}
\resizebox{16cm}{!}{
\begin{tabular}{c c c c c c c c c c c c }

$\omega [m]$ & $B [\alpha r_0^2]$ & $BII$ &  $BI$ & $G$ & $GII$ & $GI$ & $BG$ & $BIGI$ 
& $BGI$ & $BIG$ & $Total$ \\
$100$ & 2.22 & 3.20 & 4.95 & 0.40 & 1.71 & -0.34 & -0.48 & -0.46 & -3.4 & -0.62 & 5.88 \\
$1000$ & 3.19 & 0.67 & 1.46 & 0.05 & 0.36 & -0.021 & -0.07 & 0.07 & -1.0 & -0.07 & 1.45\\
$10000$ & 3.46 & 0.11 & 0.30 & 0.0057 & 0.065 & -0.0016 & -0.0076 & 0.0077 & -0.22 & -0.008 & 0.25\\

\end{tabular}
}
\caption{Contribution of the matrix terms respect to the Borsellino term (in percent) for three 
gamma ray energies}

\end{table}

\begin{figure}
  \centering
  \includegraphics[scale=0.8]{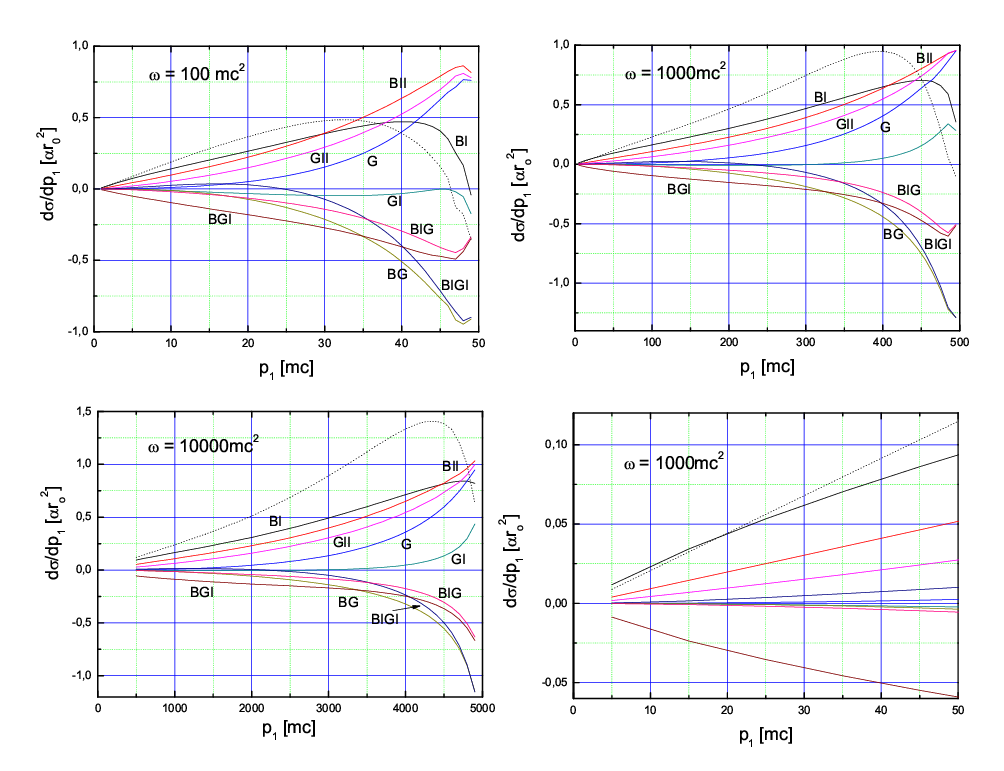}\\
  \caption{Relation between the Borsellino diagrams and the other terms. The dot line is the sum of 
all contribution. }\label{Figure 5}
\end{figure}

A better idea about the relation between the Borsellino diagrams and the other terms can be carried 
out by showing the relation $(d \sigma_B/d \rm p_1)/(d \sigma_X/d \rm p_1)$ (where $X$ represents 
any of the other terms). The results are shown in figure \ref{Figure 5} for three different $\omega$ 
and an enlargement in the $\rm p_1$ scale for $\omega = 1000m$. In this graph, it is possible to see 
that for lower value of $\rm p_1$ the curves are seen as a linear function of $\rm p_1$ (see 
\ref{Figure 5}d). Another interesting behavior that one can see is that the axes change very little 
whereas the abscissa is multiplied by 100, this indicates that the slope in the linear part is 
proportional to $\omega^{-1}$ 

Taking into account that the interval of $\rm p_1$ where the probability is not so small, and does 
not extend to where $\omega$ increases; the zone where the slope is proportional to $\omega^{-1}$ 
does not depend on $\omega$

In this way, the total correction for energy distribution obtained through the Borsellino terms can 
be written as, with an accuracy of 1\%,  $d \sigma_{tot}/d \rm p_1 = d \sigma_B/d \rm p_1$ when $\rm 
p_1 \leq m$ and for $m \leq \rm p_1 \leq 10m$, so we propose:

\begin{equation}
 \frac{d \sigma_{tot}/d \rm p_1}{d \sigma_B/d \rm p_1} = \frac{2.36}{\omega} (\rm p_1 - m)
\end{equation}

In conclusion, the energy distribution for the lower energetic electron can be expressed as ($m 
\leq \rm p_1 \leq 10m$):

\begin{equation}
 \frac{d \sigma_{tot}}{d \rm p_1} = \frac{d \sigma_B}{d \rm p_1} \left( 1 + \frac{2.36}{\omega} (\rm p_1 - m) \right) 
\end{equation}

The next step is to find a handle expression for $d \sigma_B /d \rm p_1$, in this sense one can 
find in \cite{[Boldy]} an analytical expression for $d \sigma_{Boldyshev} /d X d \Delta^2$ (eq. 37) 
where $X = 2 m (E_1  -m)$. In the same work, eq. 47 give the integration of eq. 37 over $\Delta^2$ 
and, with a simple change of variables, can be written in terms of $\rm p_1$. The eq. 47 is not very 
short and it includes the Euler's logarithm but more handled that our numerical integration.

\subsection{Energy distribution for the more energetic electron}

\begin{figure}
  \centering
  \includegraphics[scale=0.9]{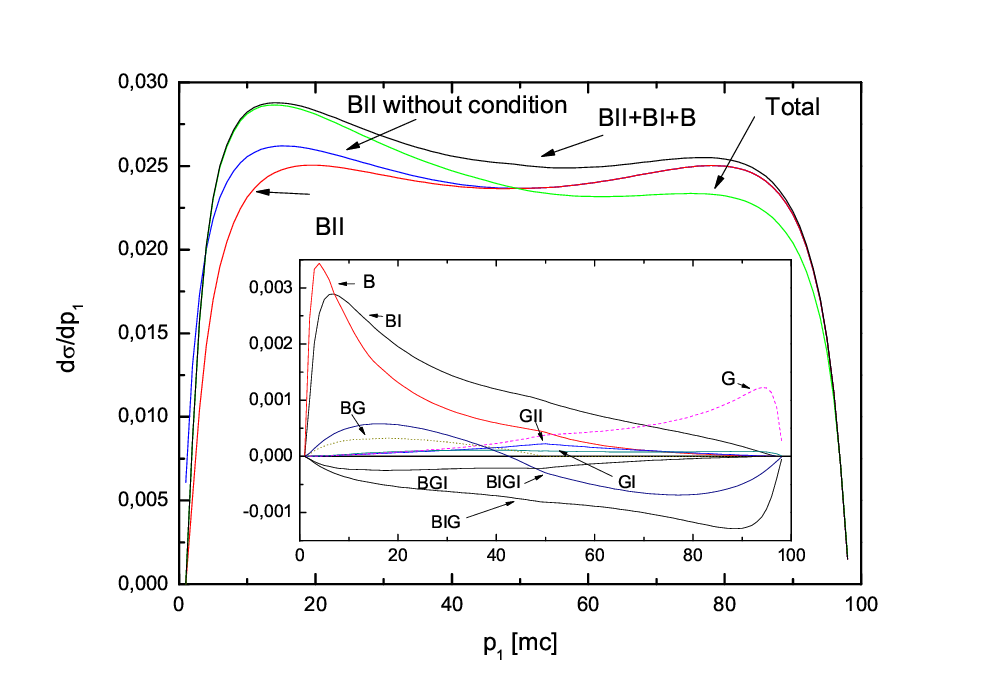}\\
  \caption{More energetic electron energy distribution, contribution from different terms. }\label{Figure 7}
\end{figure}

Now we present the results of the integration over $\rm p_2$ with the condition that it should be 
greater than a threshold detection $\rm q_0$ and lower than $\rm p_1$.
In figure \ref{Figure 7} we compare the obtained distribution taking into account different terms 
of the matrix for $\omega=100m$ and $\rm q_0=1m$. In this graph, one can see that the exchange 
Borsellino diagrams is dominant and give a curve very symmetric similar to the nuclear case (it can 
be shown that the $BII$ is similar to the Bethe-Heitler distribution and that for $\omega = 
10000m$ it can be indistinguishable between them. In the same figure, we show the curve obtained 
for the same term without any condition over the momentum; in this case the apparent symmetry is 
lost and one have more probability in the range lower than $\omega/2$. We also show the other terms, 
the total (sum of all 10 terms, which represents that one measure in an experiment taken as the 
most energetic electron like the created electron) and the sum of $BII+BI+B$. One can see that 
$BII+BI+B$ are the principal contribution but it overestimates the total distribution after the 
first peak. For $\omega=1000m$ the difference between this partial sum and the total is $1.5\%$ 
approximately (it contains more than the 90\% of the case).
In this case, the main term is $BII$ and one can repeat the same table 1 but, as it is expected, 
one can obtain the same values by making the following exchange in the notation $B \leftrightarrow 
BII$, $G \leftrightarrow GII$, $BG \leftrightarrow BIGI$ and $BIG \leftrightarrow BGI$

\begin{figure}
  \centering
  \includegraphics[scale=0.9]{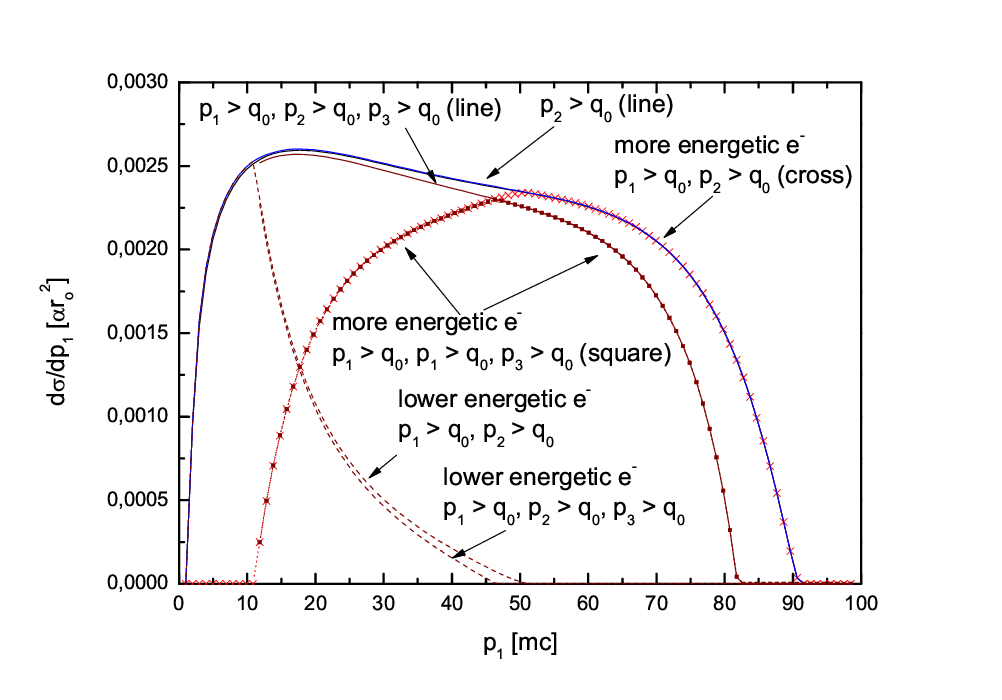}\\
  \caption{Momentum distribution for $\omega=100m$ and threshold $q_0=10m$ without restriction on 
the electrons energies (solid line), more energetic electron (dot line) and lower energetic electron 
(dash line).}\label{Figure q0}
\end{figure}

To analyze the effect of the threshold detection, we have used the $BII$ term in a extreme case, 
$\rm q_0=10m$, as an example since in this term, $\rm p_1$ is the created electron and $\rm p_2$ is the 
recoil electron, and this situation has been assumed in the literature. In figure \ref{Figure q0} we 
compare the momentum distribution obtained for $\omega=100m$ with the only condition that $\rm p_2 
\geq q_0$ (solid line), the condition $\rm p_2 \geq q_0$ and $\rm p_2 < p_1$ (more energetic 
electron, dot line) and $\rm p_2 \geq q_0$ with $\rm p_2 > p_1$ (lower energetic electron, dash 
line). The area of the first one is $0.109 \rm mb$ in agreement with the value obtained by Boldyshev et 
al (table V in \cite{[Boldy]}).

Since in the solid curve we only put the threshold on $\rm p_2$, one can see that the $\rm p_1$ 
distribution has the same form than any threshold (figure \ref{Figure 4}) in the lower range of $\rm 
p_1$ and decreases smoothly until $\geq \omega -q_0$ where it becomes zero ($\rm p_1 \geq \omega 
-q_0  \Rightarrow \rm p_2 < q_0$).

For the most energetic electron (dot curve) one can see that in the first part it is zero until 
$\rm p_1 > q_0$ since the condition $\rm p_2 < \rm p_1$ implies we do not have any electron with 
$\rm p_1 \leq q_0$. For $\rm p_1 \geq \omega -q_0 $ it is also zero due to the same reason: the 
solid curve.

The lower energetic electrons (dash curve) completed, when is added with the more energetic 
electron, the total momentum distribution. This distribution, without the condition of a threshold 
for $\rm p_2$, show a high peak at $p_1 \simeq 3m$.

If one put the threshold over $\rm p_1$ the effect is a cut at $\rm p_1=q_0$ that is $d \sigma /d 
\rm p_1 = 0$ for $\rm p_1 <q_0$ 

The positron distribution is obtained in the same way without any condition over the other 
electrons energies and not using the exchange terms ($B$ and $BII$ produce the same results and they 
do so with the other case). As was mentioned previously, Haug \cite{Haug_e+} obtained a close formula for $d \sigma = d \sigma(\rm p_3,  \theta_3)$. This formula include all the diagrams. The figure 13 in the reference \cite{Haug_e+}, show the positron energy distinction for $\omega = 100m$; we reproduce this distribution calculating all the diagrams (6 in this case) and adding them up. From this figure one can see that, contrary to electron, the position has an average energy greater than that of the electron and for $\omega\gtrsim200m$ is greater than $\omega/2$, that is, for not so high gamma ray energies, the positron, take more than the half of gamma ray energy, in average.
The Haug formula have a maximum that it is easy to parameterize according to the energy of the gamma ray so it can be write a pdf and using the rejection technique one can be sampled, taking into account the kinematics constraints, the energy and polar angle of the positron.

It is reasonable to think that the threshold detection is for the three particles, in figure 
\ref{Figure q0} we show the momentum distribution for the lower and more energetic electron and the 
total momentum distribution when the threshold detection if for the three particles. The main change 
is in the high range of $\rm p_1$ since the distribution reaches $\omega -2 q_0$ because we now have 
two particles that necessarily take momentum over $q_0$. The total cross section for this case is 
$\sigma(\rm p_1 \geq $$ 10m,\rm p_2 \geq $$ 10m,\rm p_3 \geq $$ 10m) = 0.0533 \rm mb$. 

The figure \ref{Figure q0} is an extreme case with the objective to make visible the effect; in 
a more realistic case, $\rm q_0 << \omega$, the effect of the threshold can be neglected in the most 
energetic electrons, and for the lower energetic electron it cuts the events that cannot be detected, modifying only the total cross section, for high gammas-rays energies , the results of $\sigma(\rm q 
\geq \rm q_0)$ in \cite{[Ipa]} and part of the table 5 in \cite{[Boldy]} are valid.

\begin{figure}
  \centering
  \includegraphics[scale=0.7]{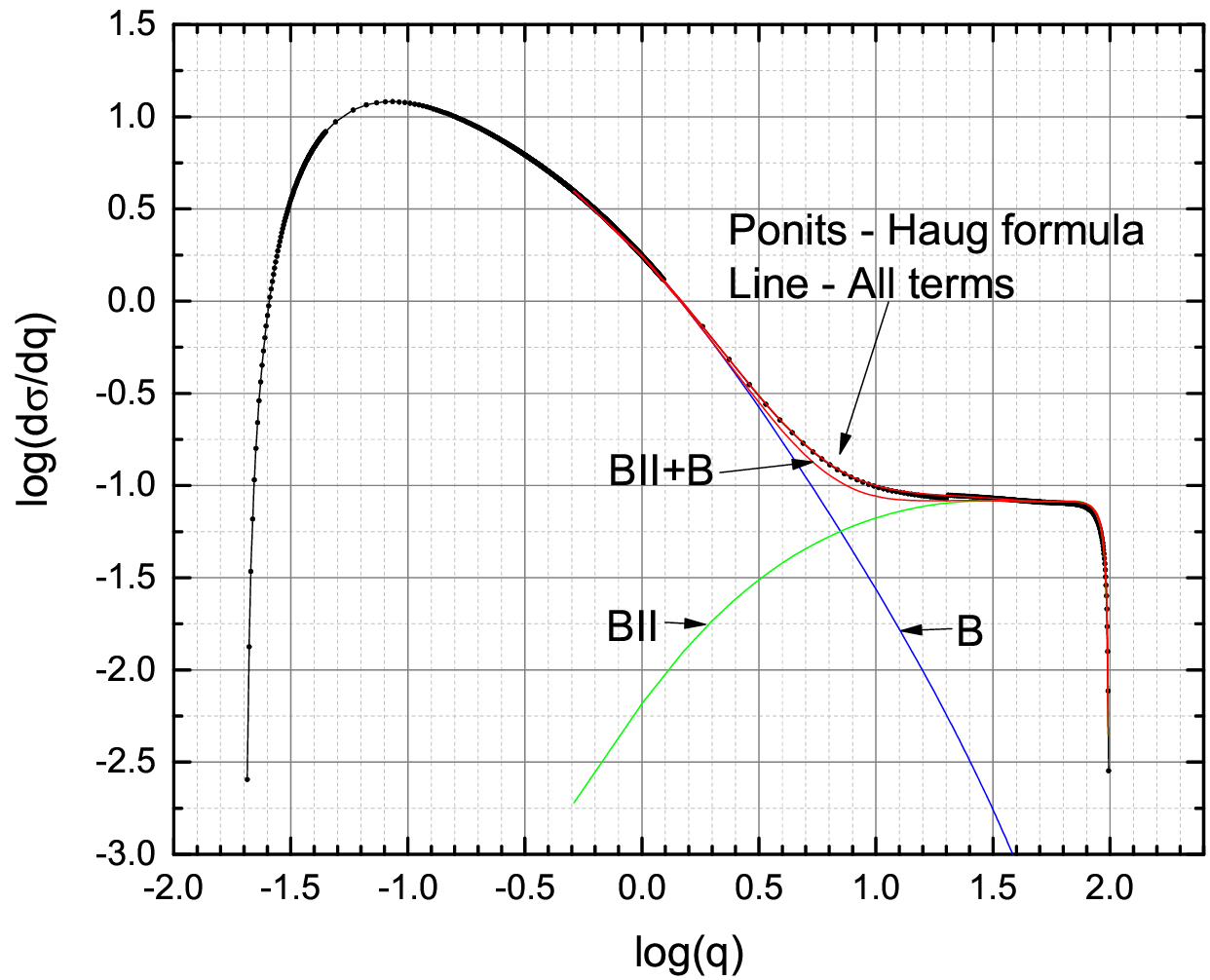}\\
  \caption{Comparison between Haug formula \cite{Haug_e-} for electron and our B, BII, B+BII and the sum of the all terms for $\omega=100m$.}\label{Haug_comp}
\end{figure}

In figure \ref{Haug_comp} we compare the Haug formula (figure 5 in reference \cite{Haug_e-}) that include the two electrons with our calculation, in this figure one can see how each term contribute to the total momentum distribution.

\begin{figure}
  \centering
  \includegraphics[scale=0.9]{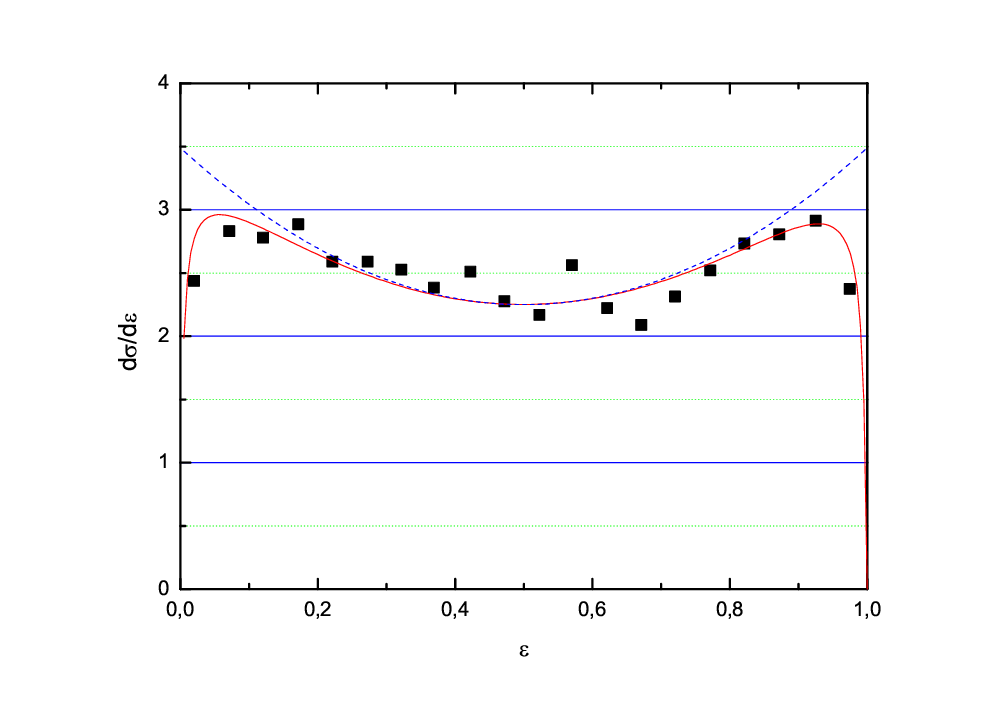}\\
  \caption{Comparison between experimental data \cite{[Ansoger]} and theoretical calculation for 
$\omega = 1000m$ and $q_0 = 1 m$.}\label{Figure 8}
\end{figure}

In the literature one can find measurements of the energy distribution, for example \cite{[GKS]}, 
\cite{[HCCS]}, \cite{[Ansoger]}. Ansoger et al \cite{[Ansoger]} shows the energy distribution in 
function of $\alpha = (E_+ - E_-)/(E_+ + E_-)$ for 4935 triplets events obtained for gamma-rays 
energies between $500-1000 \rm MeV$ and we compare this experimental result with the asymptotic 
expression obtained by Wheeler and Lamb \cite{[WL]} (for $\rm q_0=1m$) that is very similar to that one 
obtained by Boldyshev \cite{[Boldy]}. In figure \ref{Figure 8} we reproduce the figure 4 of 
\cite{[Ansoger]} and we compare it with our distribution obtained for $\omega = 1000m$; to make the 
comparison, we approximate $\alpha$ parameter of Ansoger as $(E_+ + E_-) = \omega$ so $\varepsilon 
=(\alpha - 1)/2$.
 
\section{Angular distribution}

 This section will analyses the angular distribution of triplets produced by very high energy gamma 
rays. In the previous sections, it was shown that for this regime only Borsellino diagrams 
predominate and therefore we will only use these diagrams in this analysis.     

\begin{figure}
  \centering
  \includegraphics[scale=0.7]{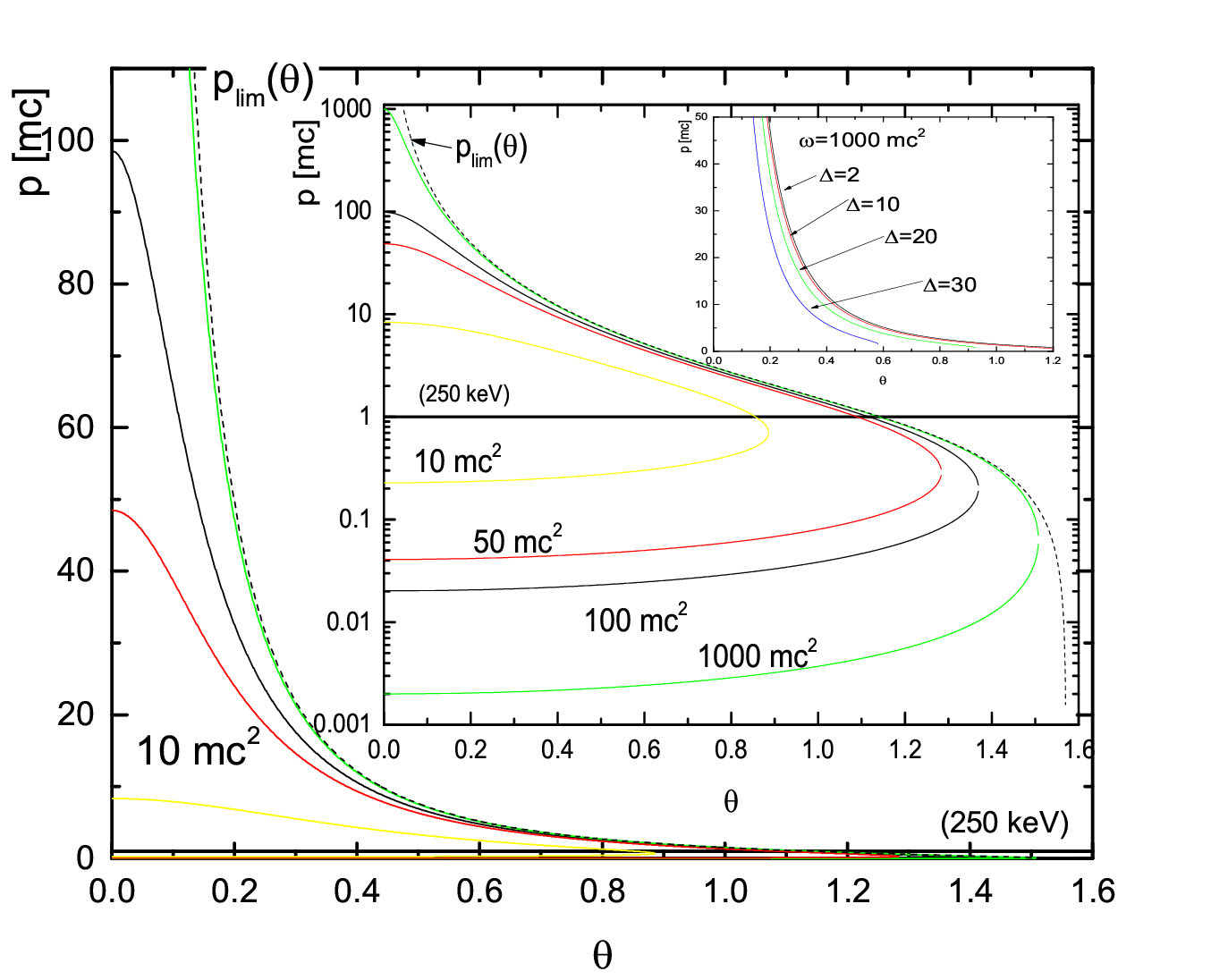}\\
  \caption{Momentum versus $\theta$ for $\Delta=2m$ and $\omega=10, 50, 100, 1000 
m$. Linear scale main graph, log scale inside the graph. Upper right corner: $p$ vs. $\theta$ for 
$\omega=1000 m$ and for $\Delta=2, 10, 20, 30 m$. }\label{Figure 10}
\end{figure}

The kinematics condition between momentum, the polar angle and the invariant mass, 
express through the relation $p(\theta,\Delta)$ (eq. 14 \cite{[Boldy]}) play an important role. In 
figure \ref{Figure 10}, we plot $p(\theta,\Delta = 2m)$ in logarithmic scale, (we choose $\Delta = 
2m$ because 
the great probability of conversion is for values of very near to $2m$, equal $2m$ means pair 
with same energies and solid angles) and decrease quickly for values larger than $2m$. We also 
show a line that represents the $p = 1m$ value to visualize how a threshold detection can affect 
the maximum angle.

From this figure one can see that large $\rm p$ (in general corresponds to pair created) have small 
$\theta$, so the transverse component of the momentum can be write as $\rm p_{i\bot} = \rm p_i \sin 
\theta_{i} \simeq \rm p_i \theta_{i}$ ($i = 1,3$). 

We will show that the angular distribution for created pair have a peak at $\theta_i \simeq m/ \rm p_i$ so, the transverse component it will expected to be equal to \rm $p_{i\perp} \simeq m$.

\begin{figure}
  \centering
  \includegraphics[scale=0.5]{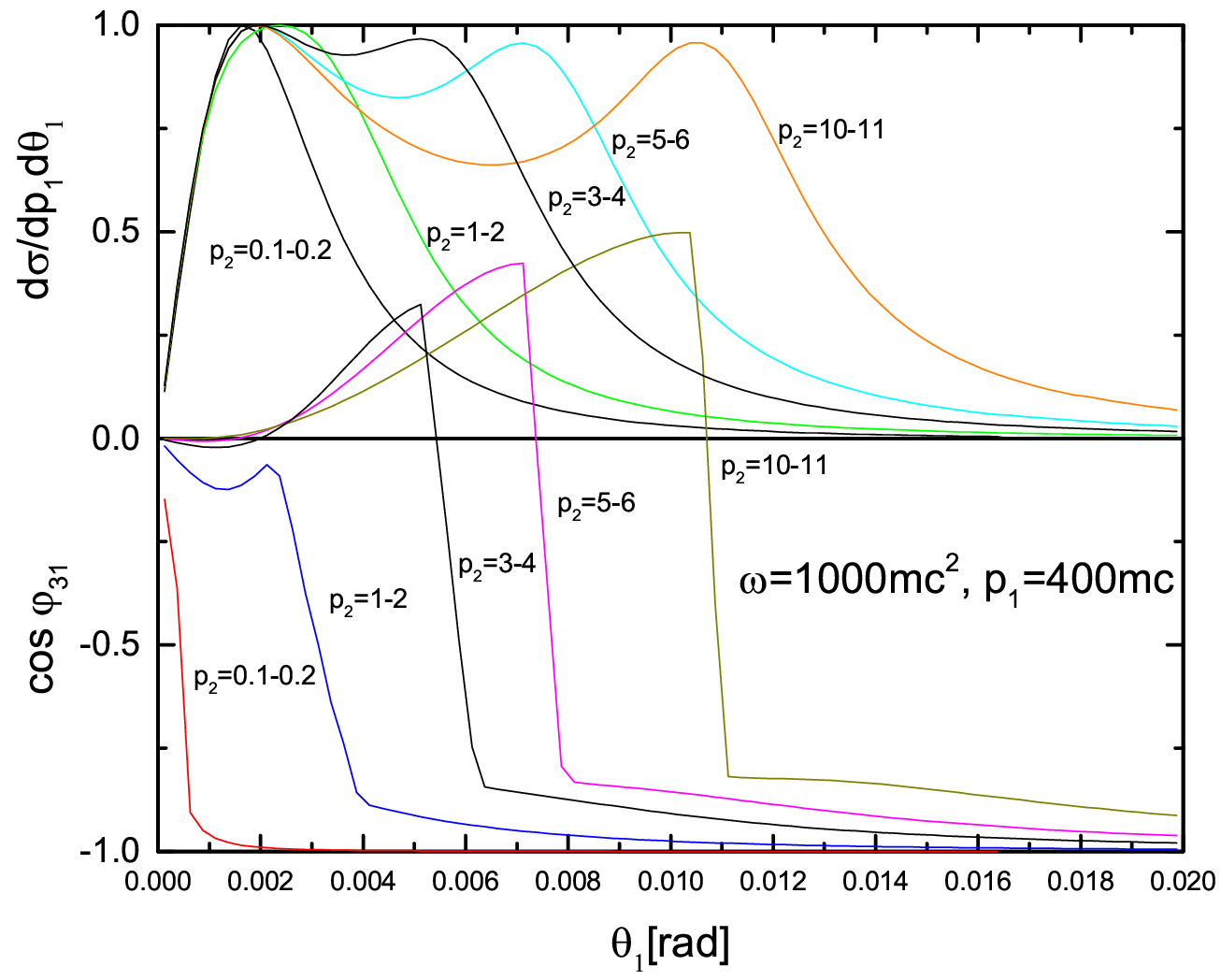}\\
  \caption{$d\sigma/dp_1d\theta_1$ and $cos\varphi_{31}$ distribution for $\omega=1000m$ and 
$p_1=400m$ }\label{Figure 11}
\end{figure}

In figure \ref{Figure 11} we plot the angular distribution for created electron (is equal for the 
positron) 
normalized to one for 
a particular value of $\rm p_1$ and different regions of $\rm p_2$. It is worth mentioning that this 
distribution is independent of the photon energy. In this figure one can see that the 
distribution has one peak when $\rm p_2 << m$ and this peak is between 

\begin{equation}
 \theta_{1,3 max-prob} \simeq \frac{0.6m}{\rm p_i}  \longrightarrow  \frac{0.7m}{\rm p_i}
\end{equation}

\noindent
and , as a consequence, $\rm p_{1,3 \perp} \simeq 0.6m$ to  $0.7m$, that is, its is approximately 
constant 
and considering that the highest probability is that $\rm p_2$ is much less than $m$, this implies that 
$\rm p_1$ and $\rm p_3$ should be approximately perpendicular to $\rm p_2$.

Integrating over all possible $\rm p_1$ values, one finds that the relationship between the maximum 
polar angle and the energy is $\theta_{max} \simeq m/\omega$ as previously reported.

\begin{figure}
  \centering
  \includegraphics[scale=0.6]{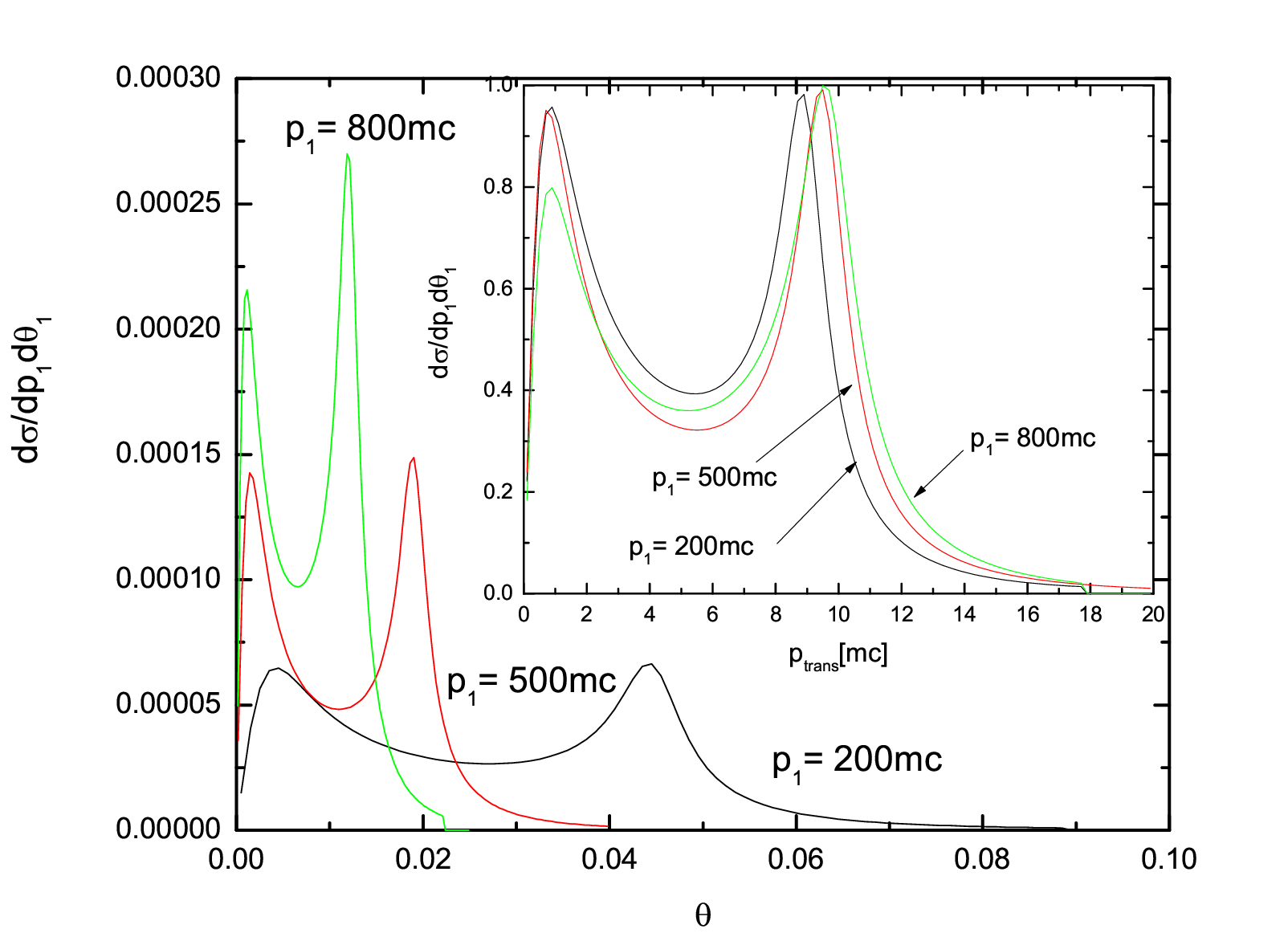}\\
  \caption{$d\sigma/dp_1d\theta_1$ and transverse momentum plotted for $\omega = 1000m$, $\rm p_2 = 
50m$ and three values of $\rm p_1$. }\label{Figure 12}
\end{figure}

For $\rm p_2 >> m$, the angular distribution shows two peaks, one coinciding with the case of $\rm p_2 
<< m$; figure \ref{Figure 12} allows us to understand the other peak where the transverse momentum 
is plotted for 
$\omega = 1000m$, $\rm p_2 = 50m$ and three values of $\rm p_1$. One can see that the peaks do not depend on $\rm p_1$ and that the second peak is around the $\rm p_{1} \simeq 9m$, from figure 10 one can see that the 
maximum polar angle for this momentum is $\theta_2 \simeq 0.2$ which implies that $\rm p_{2 \perp} \simeq  10m$.

\begin{figure}
  \centering
  \includegraphics[scale=0.55]{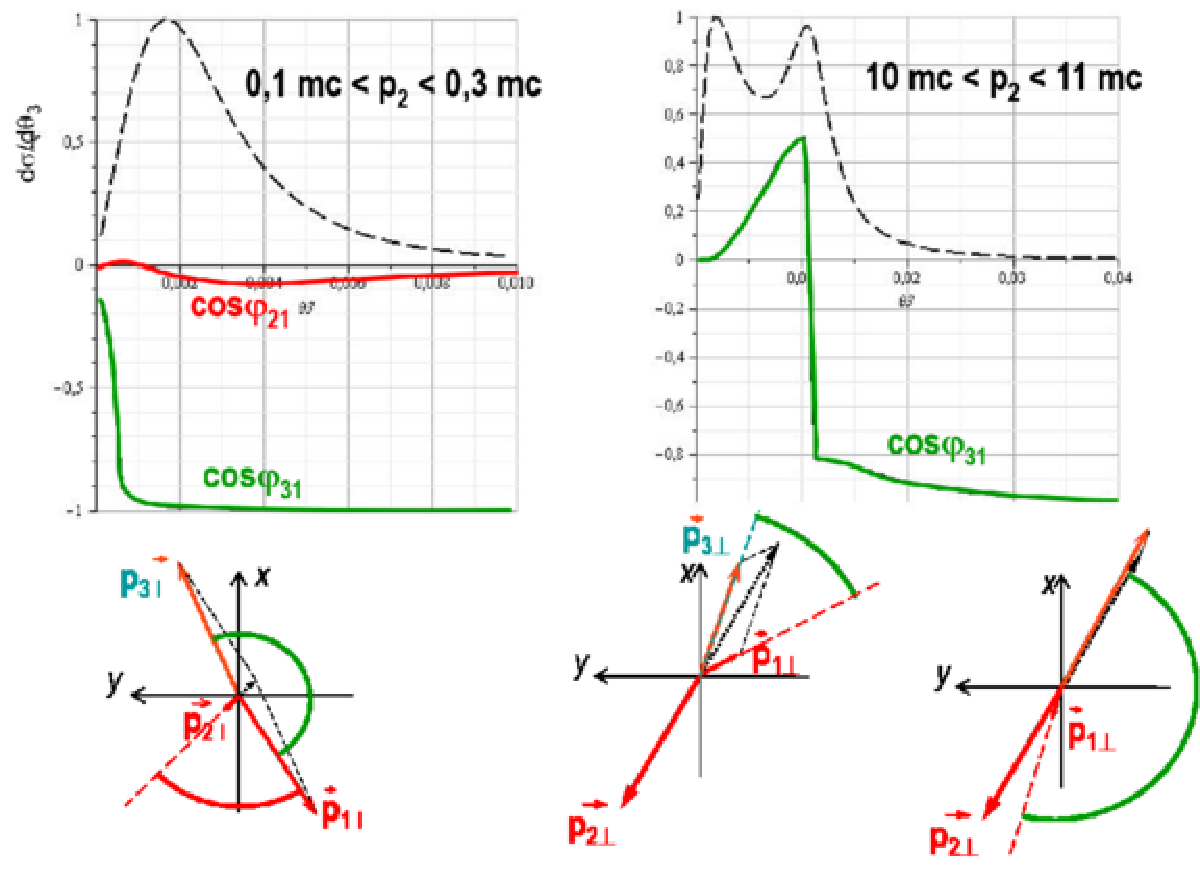}\\
  \caption{Distribution of $cos \varphi_{31}$ for two cases the $\rm p_2$ and a vectorial representation 
of the transverse momenta }\label{Figure 13}
\end{figure}

This can be interpreted as one of the particles created interacts with the electron of recoil; from 
this viewpoint, one would expect azimuthal information to be lost. To analyze this, 
figure \ref{Figure 13} graphs 
the distribution of $cos \varphi_{31}$ for two cases the $\rm p_2$; a vectorial representation of the 
transverse momenta is also shown. 

From this figure it can be seen that for $\rm p_2 < m$ (the most probable cases) the created pair is 
approximately coplanar.When $\rm p_2 >>m$, the $\varphi_{31}$ angle takes values $<\pi/2$ until $\theta_1$ 
reaches a value such that $|\rm p_{3 \perp}|$ and $|\rm p_{2 \perp}|$ are approximately equal, for 
$\theta_1$ even higher, $\varphi_{31}$ changes rapidly to values close to $\pi$. This also shows that 
$p_1$ is always much less than $m$ and that the probability of the 2 particles created interacting with the 
electron recoil is extremely low. 

\begin{figure}
  \centering
  \includegraphics[scale=0.6]{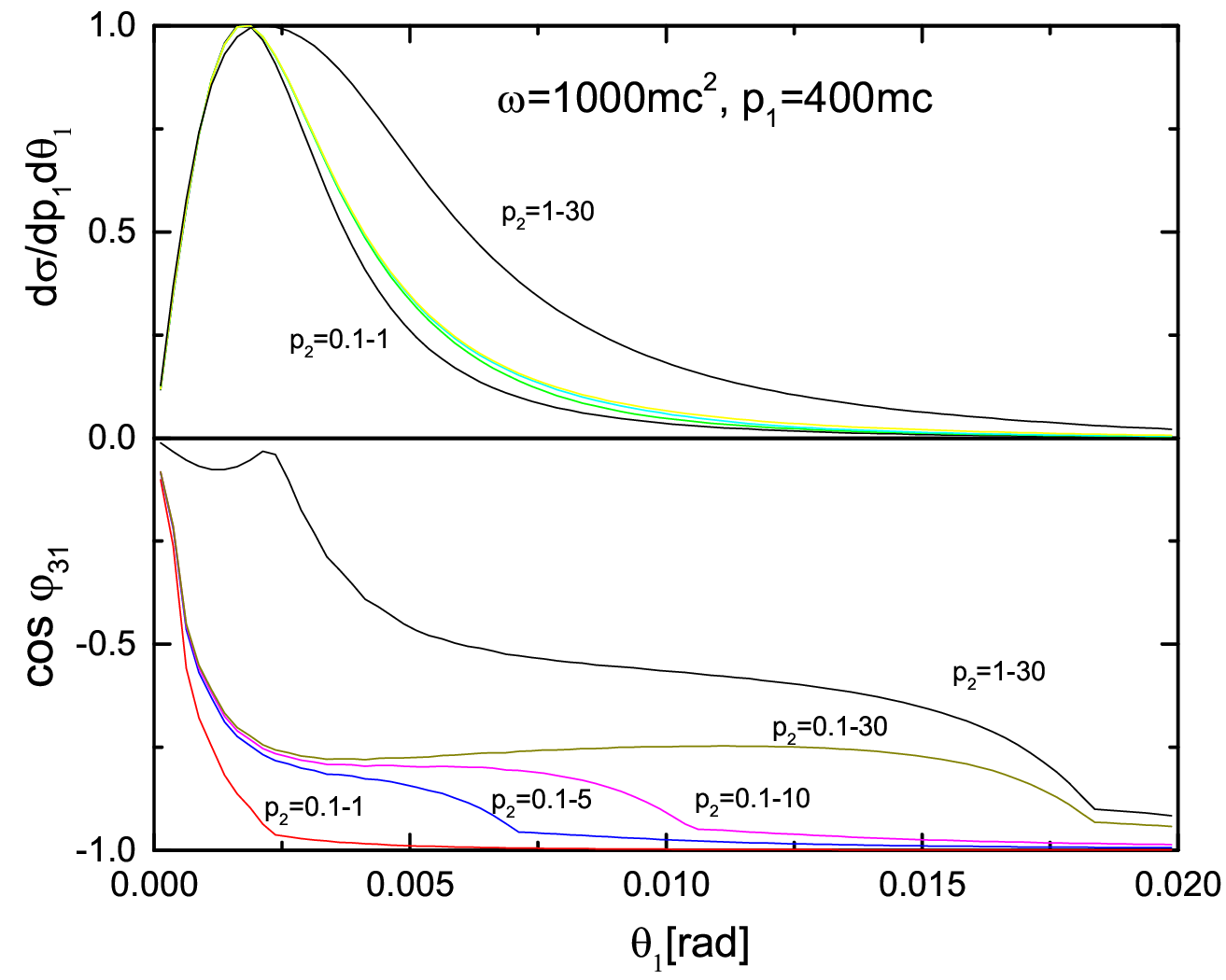}\\
  \caption{Distribution of $\theta_1$ by integrating $\rm p_2$ in different ranges indicate in the 
graph. }\label{Figure 14}
\end{figure}

Figure \ref{Figure 14} shows the distribution of $\theta_1$ by integrating $\rm p_2$ in different 
ranges. Five 
curves are shown, four of which integrate $\rm p_2$ from $0.1$ to $1,5,10,30$ respectively. The fifth 
curve is the result of integrating $\rm p_2$ between $1$ and $30$. From this figure you can see how the 
second peak is masked by the low values of $\rm p_2$ for the high probability that this process 
has. However, the probability of interaction between the recoil electron and one of the 
particles created is reflected in the distribution of $\cos \varphi_{31}$. 

\section{Asymmetry}

The cross section can always be written as:
\begin{equation}
 d\sigma = d\sigma^{(t)} (variables) [1 + P \Lambda_j (variables) cos 2\varphi_j]
\end{equation}

\noindent
where $j=1,2,3$ are any of the particles and $variables$ are polar angles and momenta of any of the 
three particles but excluding azimuthal angles and the amount goes from none to four, depending on 
the integration that one can do. In particular, through Boldyshev's \cite{[Boldy]} equations 37 
and 38, ones can graph $\Lambda (\rm p_2, \Delta)$ and observe that the asymmetry is maximum for 
$\Delta=2m$ and for low values of $\rm p_2$.

\begin{figure}
  \centering
  \includegraphics[scale=0.45]{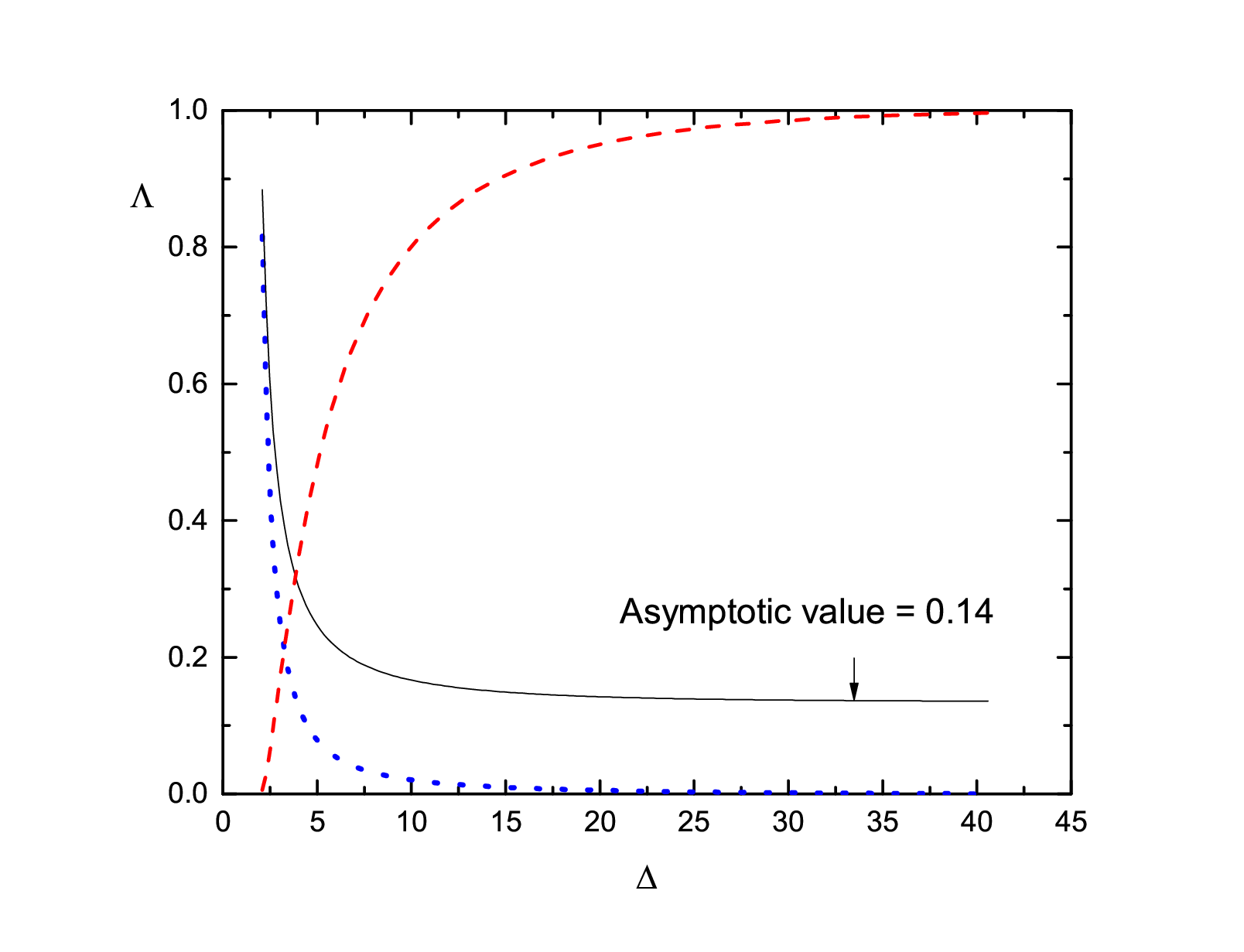}\\
  \caption{Absolute fraction of triplets as a function of $\Delta$ (solid line),
fraction of triplets produced up to the value of $\Delta$ indicated in the abscissa (dash 
line) and the asymmetry for each value of $\Delta$ (dotted line).The asymptotic value of the table 
VI of \cite{[Boldy]}is also shown.}\label{Figure 15}
\end{figure}

Figure \ref{Figure 15} shows the asymmetry as a function of the absolute fraction of triples produced 
between $\Delta_{min}$ and $\Delta$ indicated in the absise (solid
line) and the 
fraction of triplets produced up to the value of $\Delta$ indicated in the abscissa (dash line). It 
can be seen that while the asymmetry decreases rapidly when $\Delta$ moves away from the value of 
$2m$ 
(and reaches the asymptotic value of $0.14$), the fraction of triplets increases rapidly. For 
example, $50 \%$ of triplet has $2m\leq \Delta \leq 5m$ and the $80 \%$ between $2m\leq \Delta \leq 
10m$.The same figure shows the asymmetry for each value of $\Delta$ (dotted line), of this curve it 
can be seen that if only one uses the triplets with $\Delta <5m$, the asymmetry is around $25 \%$ 
but 
from figure \ref{Figure 10} figure 10 one can see how difficult it is to distinguish $\Delta$ from 
$\rm p_2$ when $\omega = 
1000m$ (making it almost impossible for $\omega = 10,000m$). 

\begin{figure}
  \centering
  \includegraphics[scale=0.45]{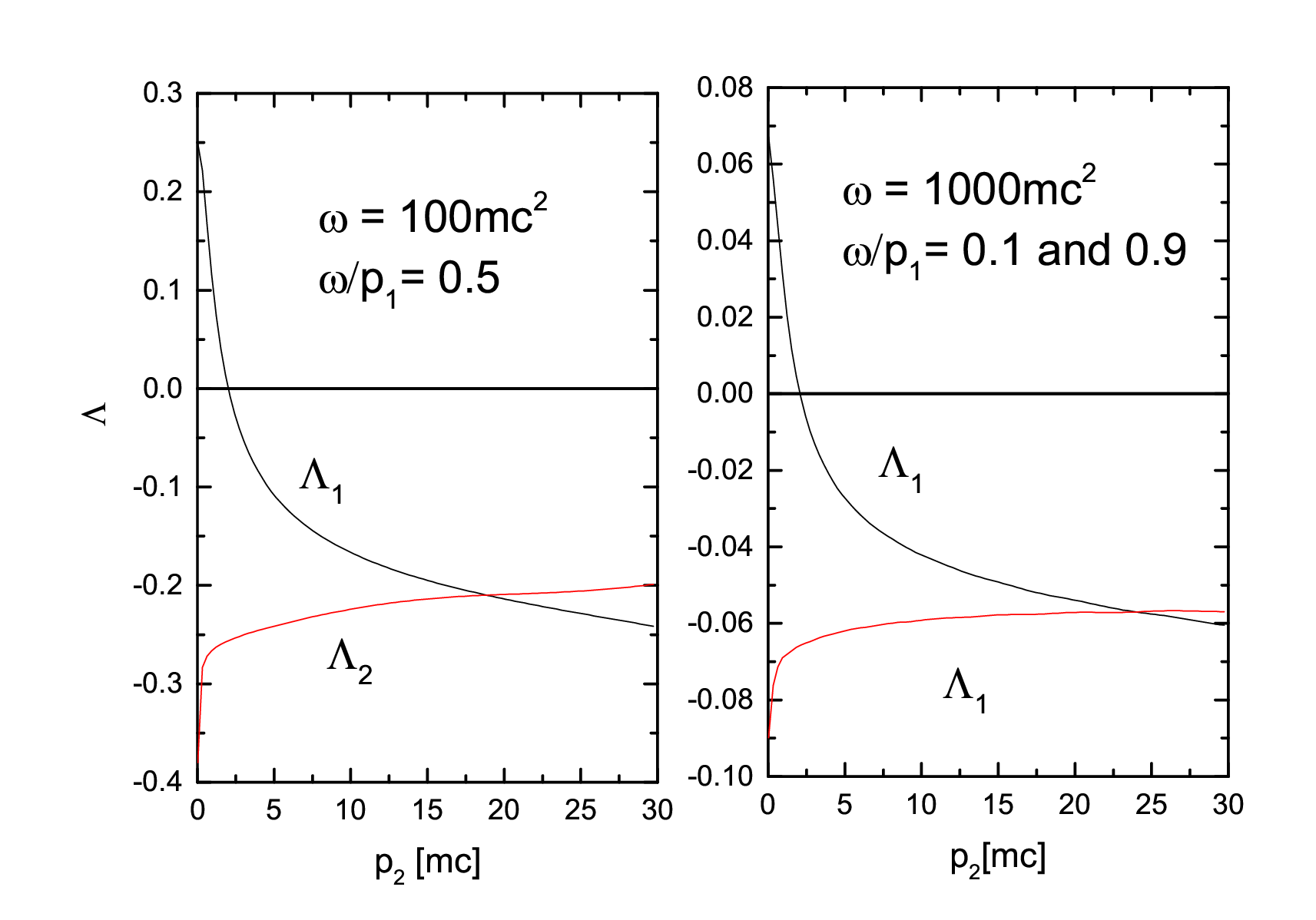}\\
  \caption{Asymmetry for one of the particles created and for the recoil electron, for some 
representative values of $\omega/\rm p_1$ as a function of $\rm p_2$. }\label{Figure 16}
\end{figure}

Symmetry analysis:

The presence of a second peak in the distribution of the polar angle for large momentum 
transferred interpreted 
as a collision between one of the created particles and the electron recoil would 
indicate that a loss of azimuthal distribution information, expecting it to become isotropic. 
Figure \ref{Figure 16} shows the asymmetry for one of the particles created and for the recoil 
electron, for some 
representative values of $\omega/p_1$ as a function of $\rm p_2$. The following conclusions can be 
drawn from the graph:

1) The $\Lambda_2$ asymmetry has a fairly constant negative value for the whole $\rm p_2$, this 
indicates that the electron recoil is placed with greater probability in perpendicular form to 
the polarization vector and it does it for any value of $\rm p_2$ and $\omega$.  

2) The fact that $\Lambda_2$ remains constant at $\rm p_2$ indicates that the classical interpretation 
of the collision between two particles and their consequent loss of azimuthal information is 
invalid.  The fact that $\Lambda_2$ remains constant for any value of $\rm p_2$ suggests that the 
azimuthal location (prior to annihilation) between polarization vector and the electron recoil  
in whose field the pair is to be produced has an important influence on the matrix elements. 

3) The remarkable variation of $\Lambda_2$ for $\rm p_2 << 0.2$ also does not support a classical 
interpretation.

4) The $\Lambda_1$ asymmetry begins with a large positive value (same order as $\Lambda_2$) for low 
$\rm p_2$ but decreases with $\rm p_2$, cancels and changes sign. This variation if it admits a classical 
interpretation by thinking that a photon traveling along the $z$ axis with polarization vector 
along the $x$ axis has a probability of creating a pair greater in the vicinity of electrons 
close to the plane $(z,y)$ and in the minimum interaction conditions ($\rm p_2$ lower) the pair is more 
likely to come out in the plane $(x, z)$ (that is, $\Lambda_1 >  0$);  by increasing the intensity 
of the interaction ($\rm p_2$ high), the particles of the pair are going to be deviated with more 
probability in the plane (y, z), this is $\Lambda_1 <0$.

\section{Conclusion}

The work is divided into two parts, on the one hand it analyzes the energy distribution of the 2 
electrons present in the final state of the event and due to the indistinguishability of the 
electrons, we have obtained the momentum distribution for 
the lower and more energetic electron instead of the recoil or created electron according to a real 
experimental situation.

We have calculated the 8 Feynman diagrams to take into account all the terms. From the 9 variables 
that the process has, we have integrate 6 in analytical form and 2 in numerical way to obtain the 
momentum distribution of one of the electrons. In the calculation, we have taken into account a 
threshold for the particle detection.

We have tested our formulas calculating the area of the distribution to obtain the total cross 
section and we have compared it with the results obtained by Joseph and Rohrlich 
\cite{[JosephRohrlich]}, Mork \cite{[Mork]} and Haug \cite{Haug_e+}. Also, we have obtained a very 
good agreement with the cross section (using only the Borsellino diagrams) as a function of the 
threshold published by Boldyshev \cite{[Boldy]} (table V).

For each distribution, we have shown the contribution of each term of the scattering matrix and 
found that the $G (GII)$ and $BIG (BGI)$ has not neglected contribution for $\omega \lesssim 1000m$ 
(at  $\omega = 1000m$ these terms have a contribution of around $ 1\%$).

For the lower energetic distribution, we have developed a simple correction of the Borsellino 
dominant term for this case, $B$ (that can be evaluated more easily using an analytical expression 
given by Boldyshev \cite{[Boldy]})

The more energetic distribution (and the distribution for the created electron) show an asymmetry 
respect to $\omega/2$ (the average energy that the created electron must be $<\omega/2$ since now 
one have another particles than the electron that must taken energy). As the lower case, $G (GII)$ 
and $BIG (BGI)$ terms cannot be neglected .

We have shown that the threshold detection for the particles, not only cuts the distributions but 
also it modifies them, but the modification is not significant when $\rm q_0 << \omega$.

Secondly, the angular distribution of particles in their final state was analyzed for the case of 
high energy (only the terms of Borsellino are used).

It was found that the transverse momentum of the particles created is approximately the same (when 
there is a low momentum transferred), regardless of the total momentum of the particles and 
independently of the gamma ray energy. 

The distribution of the polar angle shows a peak that is inversely proportional to the total momentum 
of the particle and for large momenta of transference a second peak appears which, in principle, can 
be related to an interaction between one of the created particles and the recoil electron.  

Furthermore, it was demonstrated that the azimuthal distribution of the electron of recoil is not 
affected by this interaction (as opposed to what is expected from a collision from the classical 
point of view) implying that the previous orientation between the photon (and its polarization 
vector) and the electron in whose field the pair is produced is an important factor. However, this 
hypothetical interaction between the two particles is observed in the azimuthal angle between the 
two particles created.

The asymmetry ratio is increased for pairs of particles with $\Lambda$ very close to their lower 
limit but it was shown that selecting these events from the momentum measurement of the electron 
recoil becomes very difficult for gamma rays with medium large energies and practically impossible 
for very large energies.

The constancy of the asymmetry $\Lambda_2$ and its independence with energy (for energies of high 
range rays) indicate that this is a better parameter than the asymmetry $\Lambda_1$ to determine 
polarization. 

\section*{Acknowledgment}

The authors would like to thank Prof. Denis Bernard for his invaluable comments and suggestions on 
this manuscript and Prof. Francesco Longo for their invitations to INFN di Trieste where these 
ideas were presented and discussed.

\end{document}